\documentclass[journal=ancac3,manuscript=article]{achemso}
\usepackage[version=3]{mhchem} % Formula subscripts using \ce{}
\usepackage{amsmath}
\usepackage{nameref} 
\usepackage{hyperref}

\author{Simanta Lahkar}
\affiliation[EE TUE]
{NanoComputing Research Lab, Integrated Circuits Group, Electrical Engineering Department, Eindhoven University of Technology, Eindhoven, 5612AZ, The Netherlands}

\author{Valeria Bragaglia}
\affiliation[IBM]
{IBM Research Europe-Zurich, Rueschlikon, Switzerland}
\alsoaffiliation[EE TUE]
{NanoComputing Research Lab, Integrated Circuits Group, Electrical Engineering Department, Eindhoven University of Technology, Eindhoven, 5612AZ, The Netherlands}

\author{Behnaz Bagheri}
\affiliation[AP TUE]
{Department of Applied Physics, Eindhoven University of Technology, Eindhoven, 5612AZ, The Netherlands}

\author{Donato Francesco Falcone}
\affiliation[IBM]
{IBM Research Europe-Zurich, Rueschlikon, Switzerland}

\author{Matteo Galetta}
\affiliation[IBM]
{IBM Research Europe-Zurich, Rueschlikon, Switzerland}

\author{Marilyne Sousa}
\affiliation[IBM]
{IBM Research Europe-Zurich, Rueschlikon, Switzerland}

\author{Aida Todri-Sanial}
\affiliation[EE TUE]
{NanoComputing Research Lab, Integrated Circuits Group, Electrical Engineering Department, Eindhoven University of Technology, Eindhoven, 5612AZ, The Netherlands}
\email{a.todri.sanial@tue.nl}

\title[An \textsf{achemso} demo]
    {Decoupling Electric Field and Temperature-Driven Atomistic Forming Mechanisms in TaO$_{\rm x}$/HfO$_{\rm 2}$-Based ReRAMs using Reactive Molecular Dynamics Simulations}

\begin{document}

\begin{abstract}

Resistive random access memories (ReRAMs) with a bilayer TaO$_{\rm x}$/HfO$_{\rm 2}$ stack structure have shown unique multi-level resistive switching capabilities. However, the physical processes governing their behavior, and specifically the atomistic mechanisms of forming, remain poorly understood. In this work, we present a detailed analysis of the forming mechanism at the atomic level using molecular dynamics (MD) simulations. An extended charge equilibration scheme, based on a combination of the charge transfer ionic potential (CTIP) formalism and the electrochemical dynamics with implicit degrees of freedom (EChemDID) method, is employed to model the localized effects of applied voltage. Our simulations reveal that tantalum ions exhibit the highest displacement under applied voltage, followed by hafnium ions, while oxygen ions respond only minimally. This results in the formation of a tantalum-depleted, oxygen-rich zone near the positive top electrode (anode), and the clustering of oxygen vacancies near the negative bottom electrode (cathode), where the conductive filament nucleates. This ionic segregation partially shields the bulk dielectric from the applied electric field, hindering further migration of ions in the vertical direction. We find that a minimum threshold voltage is required to initiate vacancy clustering. Filament growth proceeds through a localized mechanism, driven by thermally activated generation of oxygen vacancy defects, which are stabilized near the edge of the nucleated filament at the cathode. 
  
\end{abstract}

%%%%%%%%%%%%%%%%%%%%%%%%%%%%%%%%%%%%%%%%%%%%%%%%%%%%%%%%%%%%%%%%%%%%%
%% Start the main part of the manuscript here
%%%%%%%%%%%%%%%%%%%%%%%%%%%%%%%%%%%%%%%%%%%%%%%%%%%%%%%%%%%%%%%%%%%%%
\section{Introduction}
\label{sec:intro}

Filamentary resistive random access memories (ReRAMs) are resistive switching devices that have gained significant attention for their potential in neuromorphic computing and non-volatile memory technologies owing to their promising characteristics like deep scalability, ease of programmability, and large on/off ratios \cite{chen2012balancing,lee2011fast,govoreanu201110}. These devices are typically built using a metal-insulator-metal stack design, where the resistance of the insulator can be dynamically altered based on the history of charge flow through it. \cite{chen2012balancing,lee2011fast,dittmann2021nanoionic,ma2019stable}
However, typically, an initial process known as electroforming is required before acquiring this dynamic resistance change capability. During electroforming, a high voltage is applied across the electrodes, inducing the formation of a conductive filament within the insulating layer. Transition metal oxides (TMOs) such as \ce{TaO2}, \ce{Ta2O5}, and HfO$_{\rm 2}$ are common candidates to form the insulating layer, which are often amorphous, in conventional ReRAMs. In such TMO-based ReRAM devices, the filament is formed as a connected region of oxygen-deficient conductive phases within the dielectric, commonly pictured as a continuous network of oxygen vacancies clustered together in the insulating layer \cite{cartoixa2012transport, rosario2018correlation, rosario2019metallic, ma2018formation, urquiza2021atomistic}. The structural evolution of and around this filament allows the device to switch between low- and high-resistance states, enabling the resistive switching behavior that ReRAMs are known for \cite{chen2012balancing,lee2011fast,dittmann2021nanoionic,ma2019stable,ma2020exchange}. 

In recent years, bilayer ReRAM structures combining a layer of a conductive metal oxide (CMO) material, like TiO$_{\rm x}$, TaO$_{\rm x}$, and AlO$_{\rm x}$, and a dielectric HfO$_{\rm 2}$ layer stacked together between the electrodes have shown significant improvements over single-layer devices \cite{yao2017face,woo2016improved,cuppers2019exploiting,stecconi2022filamentary}. These bilayer configurations offer enhanced analog bipolar switching, the ability to achieve multiple resistive states, and reduced noise and stochasticity in switching. Such properties make bilayer ReRAMs particularly attractive for energy-efficient neuromorphic computing applications, such as ReRAM-based crossbar arrays for accelerating deep neural network inference and training \cite{stecconi2024analog,wan2022compute,gong2022deep,choi2025hardware}, as well as oscillatory neural networks (ONNs), where they act as tunable resistive coupling elements between oscillating neurons \cite{delacour2021oscillatory,todri2024computing,falcone2025all}.

Experimental studies on TaO$_{\rm x}$/HfO$_{\rm 2}$-based ReRAMs have shown how CMO and HfO$_{\rm 2}$ material properties impact their performance characteristics, like resistance range, switching behavior, and yield \cite{stecconi2022filamentary,stecconi2024analog}. Furthermore, a physics-based compact model \cite{galetta2024compact} accounting for the electro-thermal distribution across the material stack \cite{falcone2024analytical,falcone2023physical} has been recently proposed that can simulate their switching behavior with high accuracy. However, despite the promising characteristics of bilayer ReRAMs, the literature lacks studies addressing their electroforming process from an atomistic perspective, which is crucial to clarify and validate the physical mechanisms underlying the behavior of these devices. On the other hand, the electroforming in conventional single-layer ReRAMs has been extensively studied, providing insights into their behavior aiding in the optimization of the device characteristics \cite{cartoixa2012transport, rosario2018correlation, rosario2019metallic, ma2018formation, urquiza2021atomistic,ma2020exchange}. Nevertheless, the primary electroforming mechanism in TaO$_{\rm 2}$, \ce{Ta2O5}, or HfO$_{\rm 2}$-based ReRAMs remains a matter of contention in the literature, even for single-layer devices \cite{kumar2016conduction,kumar2016memristors,ma2018formation,ma2020exchange}.

The behavior of single-layer \ce{TaO2}, \ce{Ta2O5}, or HfO$_{\rm 2}$-based ReRAMs has conventionally been attributed to the valence change mechanism (VCM)\cite{waser2009redox}. In the VCM model, it is the electric field-driven migration of oxygen vacancies, which generate at the anode \cite{padovani2012understanding,jeong2008characteristic}, that results in the formation of a conductive filament \cite{urquiza2021atomistic,dittmann2021nanoionic,waser2009redox}. This filament can break and reform, upon subsequent reversing of the voltage polarity applied to the electrode, enabling resistive switching. Joule heating can further accelerate this process by providing the thermal energy necessary to enhance the electric field-driven migration of oxygen vacancies \cite{dittmann2021nanoionic}.

In contrast, the behavior of some ReRAMs, such as those based on NiO, is explained by the thermochemical mechanism (TCM), where thermally activated conductivity increase in the metal oxide layer creates a hot channel, leading to ionic segregation driven by the thermal gradient in the radial direction around the channel and consequent filament formation \cite{goodwill2019spontaneous,russo2009self,ielmini2011thermochemical,son2008direct}. Recent experiments on single-layer TaO$_{\rm 2}$-based ReRAM with inert electrodes have revealed a two-stage forming process, characterized by thermal runaway followed by compositional runaway, and lateral ionic segregation in the dielectric layer, which are indicative of the TCM model \cite{ma2020exchange, ma2019stable}. Other studies have also indicated that the thermal effects could be even more significant in the forming process of TaO$_{\rm 2}$, \ce{Ta2O5,} and HfO$_{\rm 2}$-based ReRAMs than the electric field-driven effects, challenging the traditional distinction between VCM and TCM in these devices \cite{goodwill2017electro,kumar2016conduction,kumar2016memristors,ma2018formation}. 

Additionally, some studies have suggested that the motion of tantalum atoms in TaO$_{\rm 2}$-based ReRAMs may be as prominent as that of the oxygen ions during filament formation \cite{ma2020exchange,ma2019stable,ma2018formation}. This is in apparent contrast with the expectation based on the diffusion constant of tantalum ions, which has been reported to be significantly lower than that of oxygen ions in the literature \cite{xiao2019comparative}. Thus, understanding the atomistic mechanisms within the TaO$_{\rm x}$ and HfO$_{\rm 2}$-based ReRAMs is crucial to reconcile these differences and to explain why they exhibit characteristics of both the VCM and TCM models. This becomes particularly important for clarifying the behavior of bilayer ReRAMs comprising both the TMO layers.

The focus of this study is to provide atomistic insights into the forming mechanisms of TaO$_{\rm x}$/HfO$_{\rm 2}$-based bilayer ReRAM device with inert electrodes, clarifying the respective roles of localized electric fields and thermal effects. To this end, we perform all-atom reactive molecular dynamics (MD) simulations using a bilayer stack model of the device informed by scanning transmission electron microscopy (STEM) and energy-dispersive X-ray spectroscopy (EDS) of experimental devices. The simulation framework combines the charge transfer ionic potential (CTIP) formalism~\cite{zhou2004modified,zhou2005charge} with the electrochemical dynamics with implicit degrees of freedom (EChemDID) method~\cite{onofrio2015voltage}, enabling simultaneous treatment of charge transfer, redox chemistry, and electrostatics under applied bias. We have implemented all the mentioned procedures in the Large-scale Atomic/Molecular Massively Parallel Simulator (LAMMPS) software~\cite{plimpton1995fast} to perform the reactive MD simulations under externally applied voltage. This combined MD simulations approach captures the complex interplay between ionic migration, electronic redistribution, and thermal effects with high spatial and temporal resolution; additional implementation details are provided in the \nameref{sec:methods} section.

We applied a 1.2 V bias to investigate early-stage electroforming, prior to the onset of thermal runaway typically observed above 4~V in similar devices \cite{stecconi2022filamentary,stecconi2024analog,ma2018formation,ma2020exchange}. This bias is sufficient to trigger filament nucleation, allowing us to examine interfacial effects and structural evolution under temperature variations. Our results show that Joule heating accelerates filament growth not by enhancing field-driven ion migration but by increasing oxygen vacancy generation and aggregation near the filament tip; consistent with experimental observations. To isolate the voltage contribution, we also simulated a lower 0.6~V bias, capturing the device response in a pre-forming regime. Comparison across these conditions reveals a threshold voltage required for initiating vacancy aggregation into a conductive filament and indicates the two distinct possible roles of temperature and applied voltage in the electroforming process of the device.

\section{Results and discussion}

\subsection{Device Layer Structure and Modeling Approach}

\begin{figure}[t]%[h!]
  \includegraphics[width=1\textwidth]{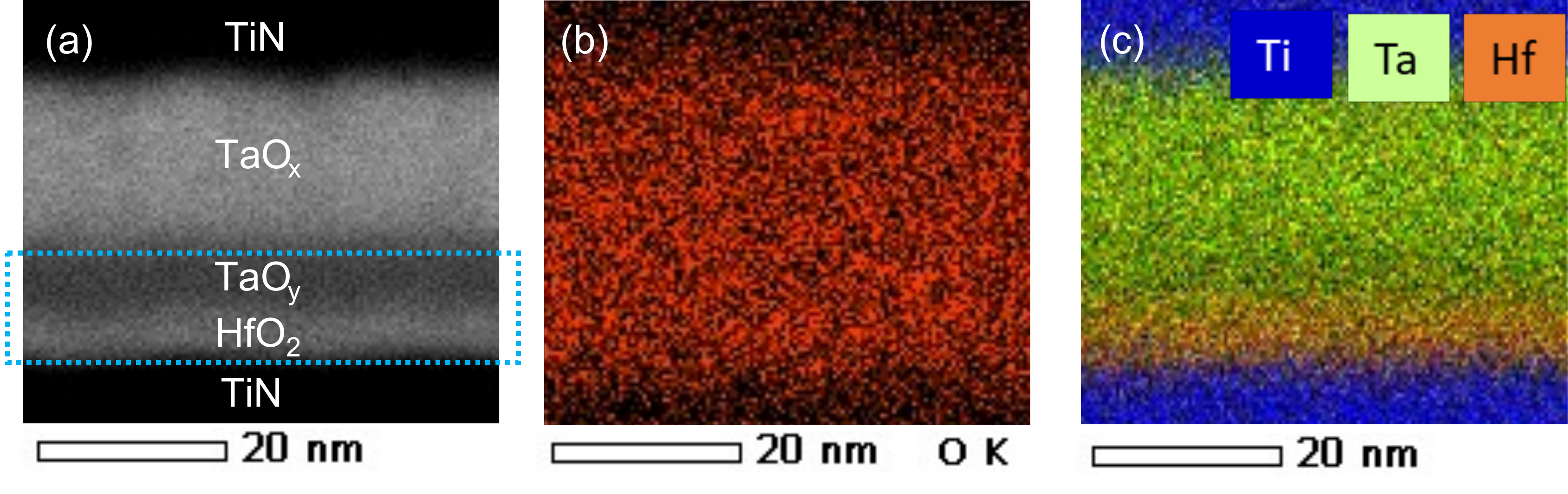}
  \caption{Cross-section images of the bilayer ReRAM. (a) STEM Bright Field image of the device cross-section. The dotted blue square highlights the two dielectric layers in the device that are oxygen-rich and amorphous. (b-c) EDS elemental maps of the device layer stack in (a).}
  \label{fgr:expt}
\end{figure}

We first obtain insights into the structure of an experimental TaO$_{\rm x}$/HfO$_{\rm 2}$-based ReRAM device, which showed multi-state analog switching behavior \cite{stecconi2022filamentary}, using transmission electron microscopy characterization to use as a reference for our atomistic model. Figure~\ref{fgr:expt} shows the scanning transmission electron microscopy (STEM) cross-section of the device and the corresponding elemental distribution maps obtained using energy dispersive X-ray spectroscopy (EDS). These devices, fabricated by depositing an oxygen-deficient conductive TaO$_{\rm x}$ layer on top of the HfO$_{\rm 2}$ layer, develop another spurious interlayer between the two, as can be seen in Figure\,\ref{fgr:expt}\,(a)\,\cite{stecconi2022filamentary,stecconi2022role}. It can be seen from the EDS elemental maps that the same elements are present in both the upper TaO$_{\rm x}$ and the spurious layer of the device, hereafter referred to as the TaO$_{\rm y}$ layer in this manuscript. An EDS line scan, shown in Figure~S1 of the Supplementary Information (SI), indicates that the TaO$_{\rm y}$ could have a higher proportion of oxygen to tantalum than the TaO$_{\rm x}$ layer. Prior characterization on the same device has shown that the TaO$_{\rm x}$ layer is conductive while the oxygen-rich TaO$_{\rm y}$ and HfO$_{\rm 2}$ layers are amorphous and dielectric \cite{stecconi2022role,stecconi2022filamentary}. For this study, we considered the composition of the oxidized TaO$_{\rm y}$ intermediate layer as that of \ce{Ta2O5}, since it is the most stable and common insulator phase of tantalum oxide\,\cite{li2019tuning}. Any electric field-assisted structural changes associated with the filament formation should initiate and occur within the regions with the highest resistivity in the device that also experience the highest potential drop during electroforming, as with the case for the region of interest for the switching behavior \cite{falcone2023physical}. Hence, we focused on the dielectric TaO$_{\rm y}$ and HfO$_{\rm 2}$ layers, hereafter referred to as the ``\textit{functional (bi)layers}'', to simulate and understand the mechanisms governing the onset and progress of the electroforming process in these devices. Additionally, both the TiN electrodes and the upper conductive TaO$_{\rm x}$ layer have low oxygen affinities, i.e., they do not have the tendency to accept or consume O ions from the dielectric matrix of the device, and, hence, are not expected to exchange oxygen ions with the functional layers \cite{stecconi2022filamentary,schonhals2018role}. Therefore, we considered the conductive upper TaO$_{\rm x}$ layer simply as an extension of the top electrode in our atomistic simulations. Furthermore, the thickness of the two dielectric layers of interest in this device (i.e., TaO$_{\rm y}$ and HfO$_{\rm 2}$) lies between 2 to 4\,nm each \cite{stecconi2022filamentary}. These insights formed the basis for creating a suitable atomistic electrochemical cell model, as described in the following two paragraphs, that can elucidate the key mechanisms governing the forming behavior of bilayer TaO$_{\rm x}$/HfO$_{\rm 2}$-based ReRAM devices.

\begin{figure}[ht]
  \includegraphics[width=0.8\textwidth]{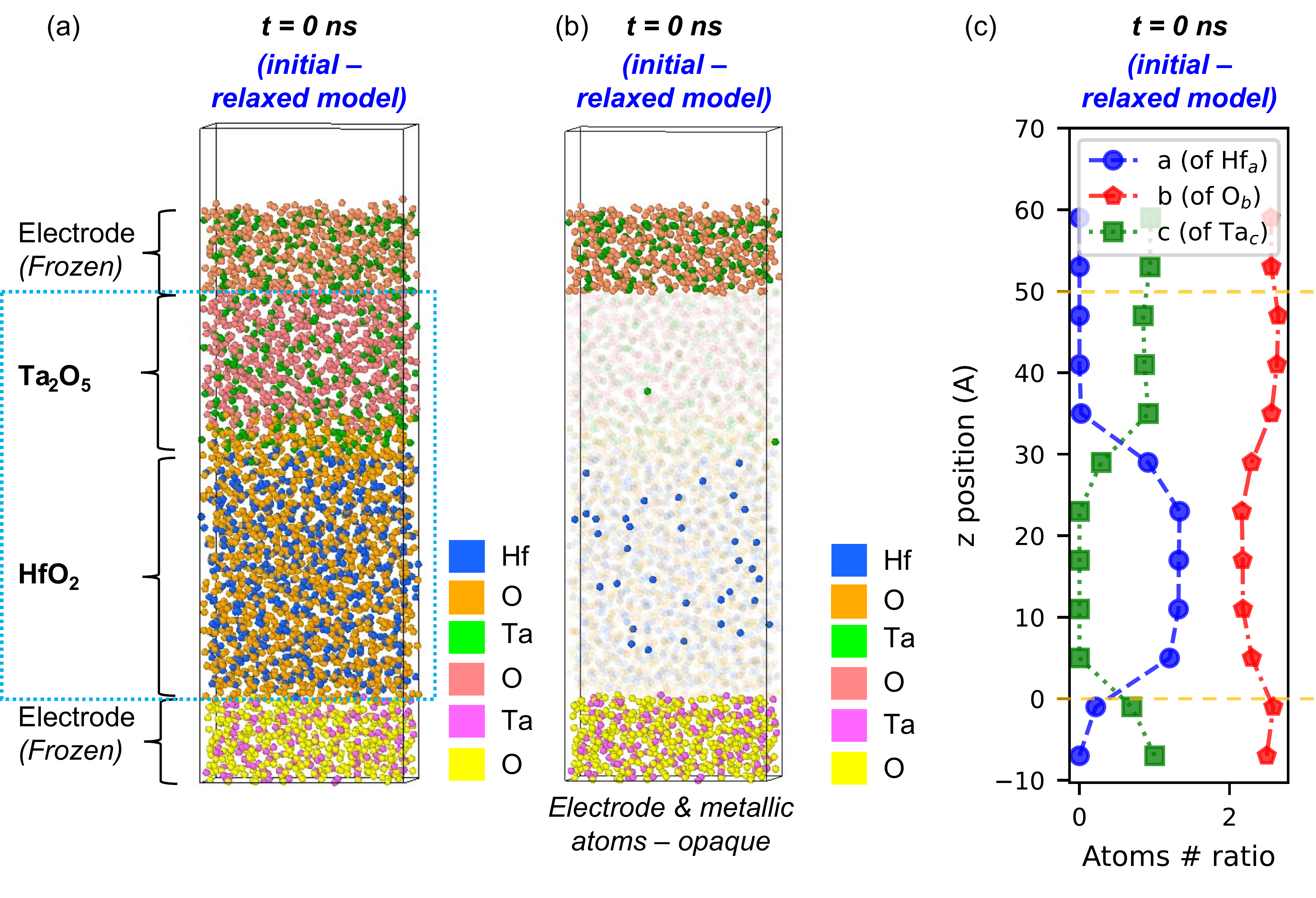}
  \caption{Structural characteristics of the relaxed \ce{Ta2O5}/\ce{HfO2} ReRAM atomistic model. (a) An atomic snapshot of the device stack model illustrating its different layers. The dotted blue square highlights the amorphous functional dielectric layers of the device, corresponding to the highlighted region of the STEM of the device cross-section Figure\,\ref{fgr:expt}. (b) Distribution of conductive (undercoordinated metal) ions between the electrodes in the initial structure, illustrated by making the rest of the atoms semi-transparent. (c) Varying material composition, given by Hf$_a$Ta$_c$O$_b$, across the thickness of the relaxed device stack model. The $a$, $b$, and $c$ values, corresponding to the proportion of Hf, O, and Ta atoms, respectively, in the structure at any position along the z-axis, are normalized so that $a+b+c$ equals 3.5. The upper and lower dotted yellow lines in the plot mark the location of the electrode/functional layer interfaces. The Tantalum and Oxygen atoms of the different layers are colored differently in the atomic snapshots for better visual distinction between the elements belonging to consecutive layers.}
  \label{fgr:model}
\end{figure}

To model the functional layers of the device, we equilibrated a stack of amorphous \ce{Ta2O5} layer on top of amorphous \ce{HfO2} layer (Figure \ref{fgr:model}\,(a)), both of which were created independently through the melt-quench process (see \ref{sec:methods} section for details). The equilibration was done for 500 ps under 1~bar pressure and at 300~K temperature in an NPT ensemble, where the number of atoms (N), pressure (P), and temperature (T) are kept constant using the Nosé-Hoover thermostat and barostat~\cite{nose1984unified,hoover1985canonical}. An upper 1\,nm thick region of \ce{Ta2O5} top layer was defined as the top electrode, leaving behind a functional bilayer of 5 nm thickness (Figure \ref{fgr:model}\,(a)). The interface of the top electrode with the rest of the \ce{Ta2O5} layer is taken to model the interface between the upper conductive TaO$_{\rm x}$ layer with the TaO$_{\rm y}$ interlayer in the actual device (Figure \ref{fgr:expt}\,(a)) \cite{stecconi2022filamentary}. The material for the bottom electrode, stacked underneath the lower \ce{HfO2} layer, was also chosen as amorphous \ce{Ta2O5} with a thickness of 1\,nm, similar to the top electrode, which allows the atomic interactions in the whole model to be defined consistently using the same interatomic potential optimized for metal oxide systems \cite{wu2023developing}. It was already clarified in a previous study that the observed characteristics of the TaO$_{\rm x}$/HfO$_{\rm 2}$-based ReRAM device cannot be attributed to any asymmetry between the work functions of the two electrodes \cite{stecconi2022filamentary}. Hence, our simulation cell, which comprises only of the two amorphous dielectric layers of interest in this device stacked between the electrodes, is a simplified electrochemical cell model that captures the key functional elements of the actual device with regards to the mechanisms of filament formation. Furthermore, the atoms in both the electrodes were kept frozen during the MD simulations under applied bias, modeling the low O affinities of both the TiN electrode and conductive TaO$_{\rm x}$ layer materials. Since the MD simulations cover a much smaller timescale as compared to the actual electroforming process, which takes several microseconds \cite{ma2020exchange}, keeping the electrodes frozen should be a sufficiently valid approximation of the effect of low oxygen affinities of the electrodes for simulating the key atomistic processes underlying the early stages of filament formation in the device.

The simulation cell for the device stack model (Figure~\ref{fgr:model}\,(a)) has a periodic boundary in x and y (lateral), and a fixed boundary in z (vertical) directions with a 1\,nm vacuum added above the top electrode. To identify and define the conductive (metallic) phases in our device stack model, whose clustering could be used to track the evolution of the filament, an oxygen coordination $\leq 5$ was adopted as the criterion for the metal atoms that would be identified as being conductive, following a previous study on \ce{HfO2}-based electrochemical cell by Urquiza \textit{et al.} \cite{urquiza2021atomistic}. The reasoning for this is based on previous first-principles studies that demonstrated conductive channels in metal/\ce{HfO2}/metal structures with filaments that comprised of 5-fold coordinated Hf atoms \cite{cartoixa2012transport}. We used the same maximum oxygen coordination criterion for both Ta and Hf atoms for simplicity and considering that the structure evolutions in the oxides of Ta and Hf with the change in O content are similar to each other \cite{xiao2019comparative}. The cluster analysis of these atoms was done using a cut-off of 3.9~\AA to identify their largest non-simply connected network, considered as the filament \cite{urquiza2021atomistic}. We find that the conductive atoms are scattered more densely in the \ce{HfO2} layer, with a more sparse distribution present in the \ce{Ta2O5} layer (Figure~\ref{fgr:model}\,(b)). This is because our adopted coordination criterion for conductive atoms is more strict with respect to the \ce{Ta2O5} layer than the \ce{HfO2} owing to the higher oxygen percentage of the former. The profiles for the proportion of different atomic species in the relaxed structure shown in Figure \ref{fgr:model}\,(c) reflect the bilayer \ce{Ta2O5} and \ce{HfO2} stack design, with a small region of intermixing at the interface between the two functional layers.

\subsection{Atomistic Response to 1.2 V applied across pristine device stack}

\begin{figure}[ht]
  \includegraphics[width=0.7\textwidth]{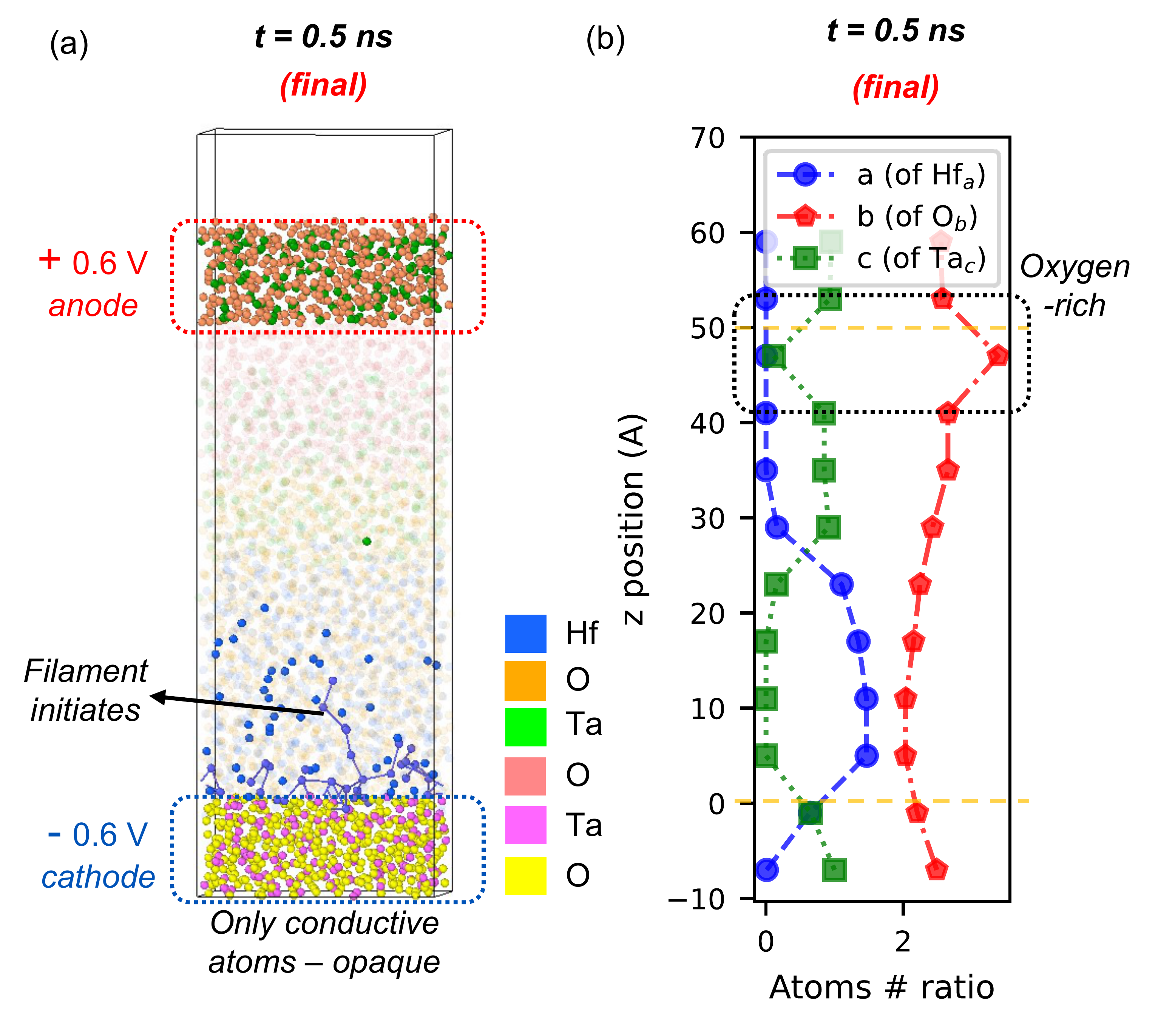}
  \caption{Structural characteristics of the bilayer stack model at the end of 500\,ps of MD simulation at 300\,K and with 1.2\,V net voltage applied across the electrodes. (a) An atomic snapshot illustrating the initialization of filament formation by clustering of conductive metal atoms (analogous to sites of oxygen vacancies) in the functional layer near the cathode. Atoms in the functional layer that are not metallic are made semi-transparent for clarity. (b) The profiles of material composition, given by Hf$_a$Ta$_c$O$_b$, along the thickness of the final device stack structure. The a, b, and c values, corresponding to the proportion of Hf, O, and Ta atoms, respectively, in the structure at any position along the z-axis, are normalized so that $a+b+c$ equals 3.5. The upper and lower dotted yellow lines in the plot mark the location of the electrode/functional layer interfaces.}
  \label{fgr:structurechange}
\end{figure}

We first studied the structural changes occurring in the device at the onset of the electroforming process. To do this, we applied a voltage of 1.2\,V across the electrodes with the top and the bottom electrodes acting as the positive (anode) and negative (cathode) electrodes, respectively, as shown in Figure \ref{fgr:structurechange}\,(a), and performed an MD simulation at 300\,K for 500\,ps in a canonical (NVT) ensemble, where the number of particles (N), volume (V), and temperature (T) were kept constant. We found that, on the application of this voltage, the scattered conductive ions in the initial structure depicted in Figure~\ref{fgr:model}\,(b) start to come closer together drifting downwards, and cluster at the cathode interface initializing the filament formation as shown in Figure~\ref{fgr:structurechange}\,(a). Additionally, a notable increase in the ratio of oxygen-to-tantalum atoms was observed on top of the \ce{Ta2O5} functional layer at its interface with the anode (Figure \ref{fgr:structurechange}\,(b)). There was also an increase in the Hf percentage in the \ce{HfO2} functional layer near its interface with the cathode, although it is not as significant a change as that near the anode (Figure \ref{fgr:structurechange}\,(b)) and can be attributed to the observed accumulation of the metallic phases (or, analogously, the oxygen vacancies) near the cathode associated with seeding of the filament (Figure~\ref{fgr:structurechange}\,(a)). In order to understand the nature and origin of these structural changes during the early stages of the electroforming process, we took a deeper look at them from the ionic perspective.

\begin{figure}[h!]
  \includegraphics[width=0.45\textwidth]{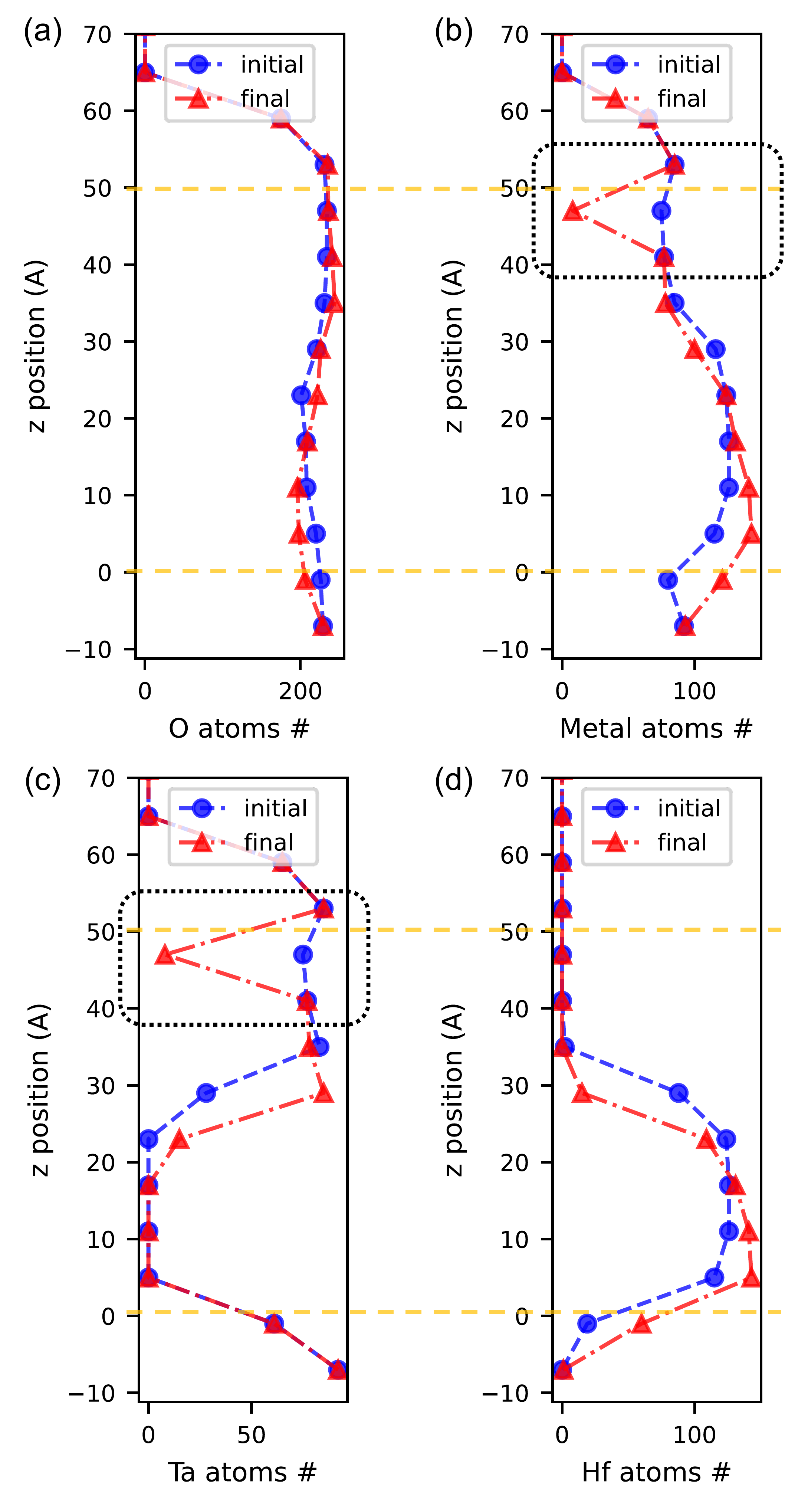}
  \caption{Change in distribution of different ionic species in the device stack over 500\,ps MD simulation at 300\,K under 1.2\,V applied voltage. The distribution profiles of (a) oxygen, (b) metal, (c) Ta, and (d) Hf atoms in the initial (relaxed) and final device stack models across their thickness (z-direction). The dotted rectangles highlight the region near the anodic interface undergoing a drastic drop in the concentration of metallic Ta ions. The upper and lower dotted yellow lines in each subplot mark the location of the electrodes/functional layer interfaces.}
  \label{fgr:ionprofiles}
\end{figure}

We calculated the distribution of each ionic species across the thickness of our device stack model in the initial relaxed state and after the 500 ps room temperature simulation under the 1.2 V applied bias, as shown in Figure \ref{fgr:ionprofiles}. We found no notable change in the distribution of oxygen atoms while a significant drop in the number of metal atoms, specifically of Ta ions, occurred near the anode (Figure \ref{fgr:ionprofiles}\,(a-c)). 

Hence, the development of the significant oxygen-rich region on top of the \ce{Ta2O5} functional layer at its interface with the anode, as noted from Figure\,\ref{fgr:structurechange}\,(b), could be attributed to a significant reduction in the Ta ion concentration, rather than an increase in the concentration of O ions near the anode, on the application of the voltage. Simultaneously, at the lower end of the \ce{HfO2} functional layer, there occurred only a minor reduction in the O ion concentration while the population of Hf ions increased much more at the cathode interface, as can be observed in Figure\,\ref{fgr:ionprofiles}\,(a and d).

\begin{figure}[h!]
  \includegraphics[width=0.5\textwidth]{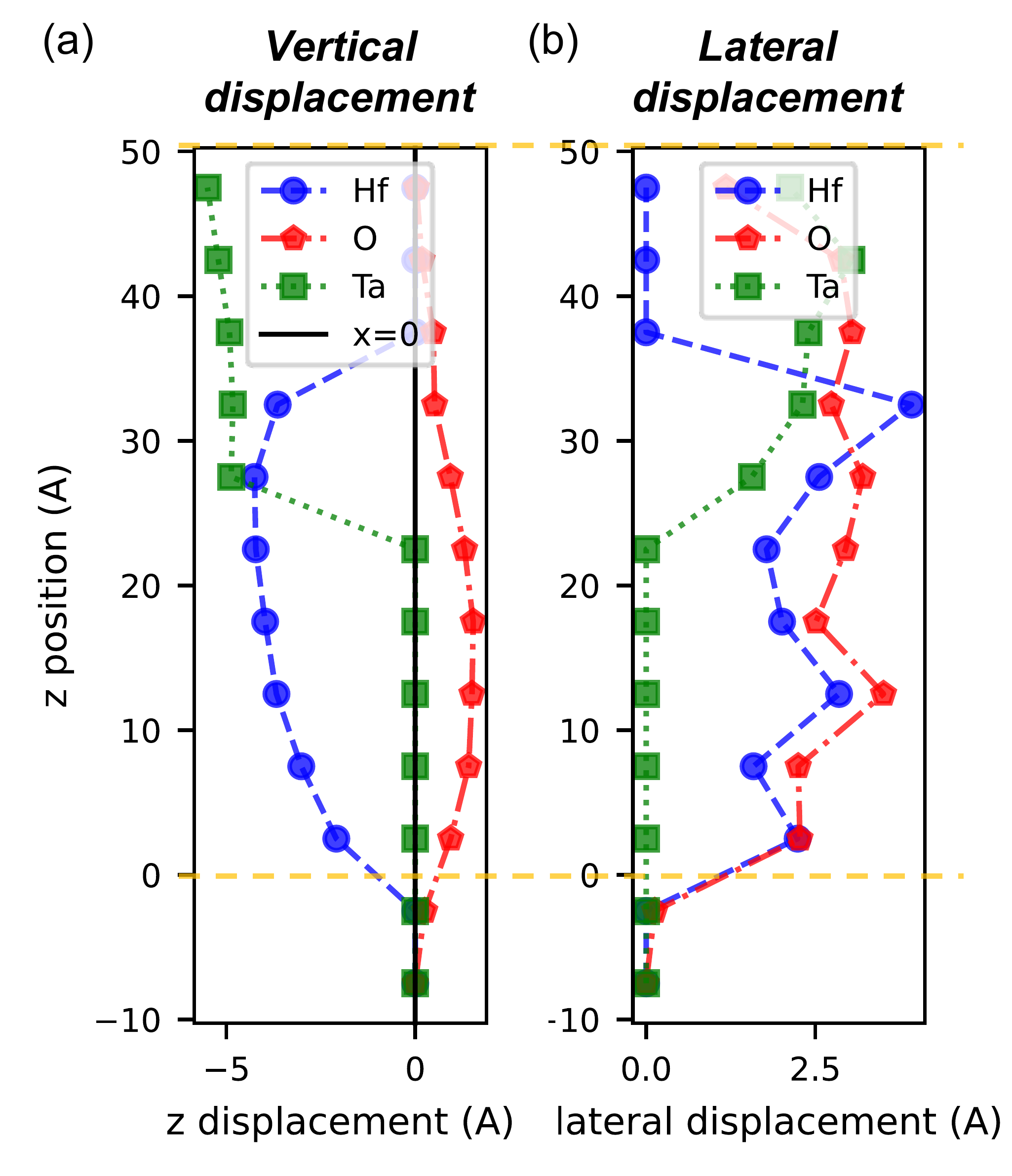}
  \caption{Variation in the motion of different ions across the device stack model under a bias of 1.2 V. The distribution profiles of net displacements of each atom type for varying initial positions along the thickness of the device stack model in (a) vertical and (b) lateral directions during 500\,ps MD simulation at 300\,K and under 1.2\,V voltage difference applied across the electrodes. The upper and lower dotted yellow lines in each subplot mark the location of the interfaces of the functional layers with the anode and the cathode, respectively, in the device model.}
  \label{fgr:ionmotion300K}
\end{figure}

Further insights into the ionic rearrangements within the device caused by the 1.2 V bias could be obtained by studying the directional displacements of each atomic species. Figure \ref{fgr:ionmotion300K}(a) shows the total voltage-driven displacements of each ionic species at different positions along the device thickness. The net displacements of the cations and anions were in the downward and upward directions, as expected from the polarity of the applied voltage (Figure \ref{fgr:ionmotion300K}(a)). We further observed that the downward displacements of the metallic ions were significantly higher than the upward displacement of the oxygen ions at any point across the device thickness (Figure \ref{fgr:ionmotion300K}(a)). In contrast, the net lateral motion, which is primarily thermally activated, was similar for all the ionic species as revealed in Figure\,\ref{fgr:ionmotion300K}\,(b). Thus, the reduction in metal ions (Figure\,\ref{fgr:ionprofiles}\,(b and c)), which created the oxygen-rich zone near the anodic interface of the \ce{Ta2O5} layer (Figure \ref{fgr:structurechange}\,(b)), was due to the Ta ions moving downwards away from the anode and into the functional layers of the device. 

These results highlight a key difference in how the ionic species respond to the applied voltage: the Ta ions at the anode experience a much stronger repulsion than the oxygen ions at the cathode, leading to larger vertical displacements for the metal cations compared to the oxygen anions. To better understand the origin of these differences, we further analyzed the charge distribution on the various atomic species within the device stack model.

\begin{figure}[!htbp]
  \includegraphics[width=0.5\textwidth]{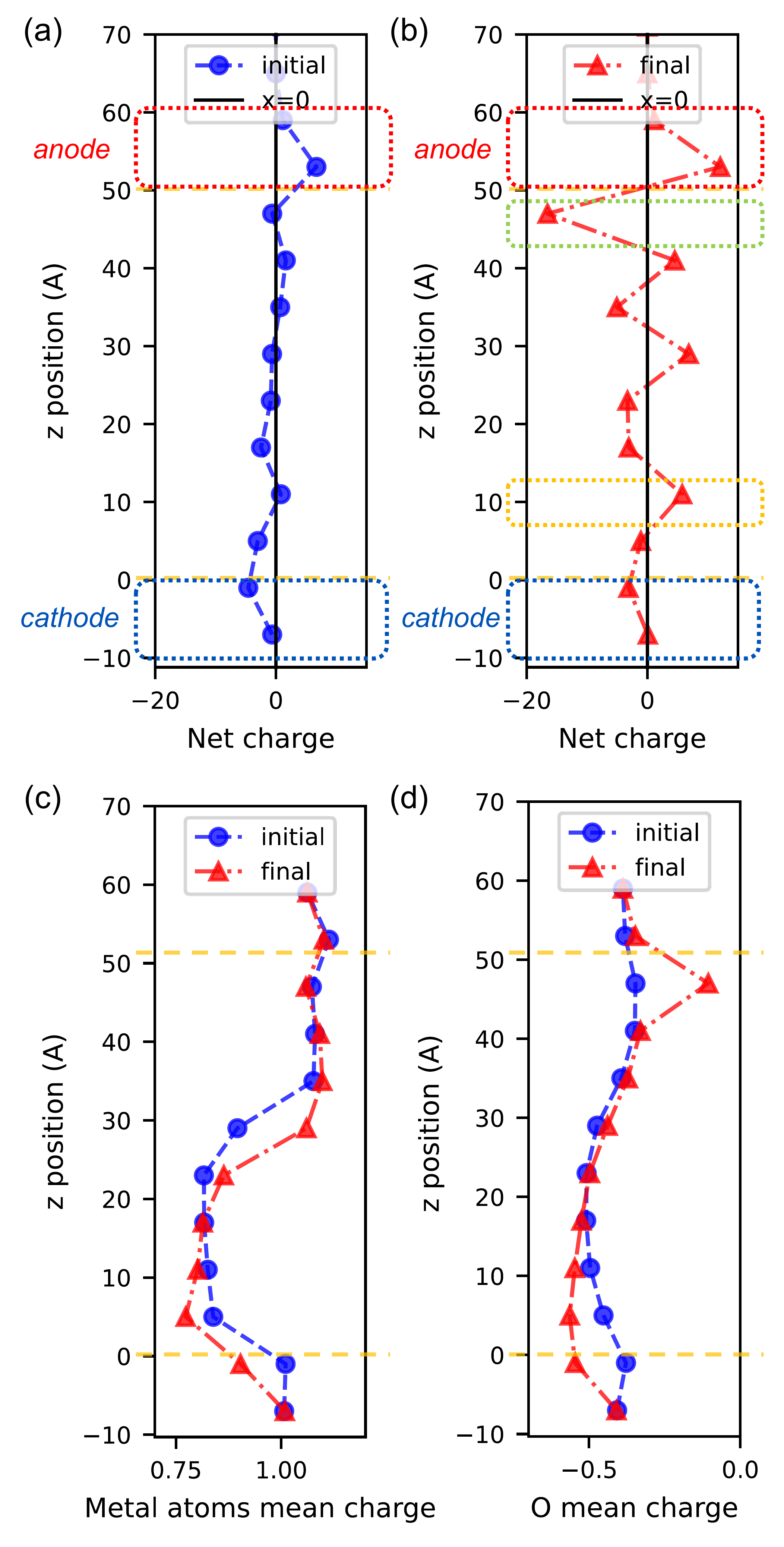}
  \caption{Distribution of ionic charges across the device stack model at the start (initial) and end (final) of 500 ps MD simulation at 300 K and under 1.2 V bias across the electrodes. The variation of net charge on all atoms across the thickness of the (a) initial (after the voltage is applied to the relaxed structure and atomic charges have been calculated) and (b) final device stack model. (c) and (d) plot of the distributions of the mean charges on the metal and oxygen ions, respectively, in the initial and final device models. The upper and lower dotted yellow lines mark the location of the interfaces of the functional layers with the anode and the cathode, respectively, in the device model. The red and blue dotted truncated rectangles in both (a) and (b) highlight the net positive and negative charges accumulated in the anode and cathode, respectively, while the green and orange truncated rectangles in (b) highlight the accumulation of net negative and positive charges, respectively, within the dielectric layers adjacent to the corresponding electrode interfaces.}
  \label{fgr:chargedistribution}
\end{figure}

We first looked at the distribution of total charge across the thickness of the device model at the start and end of the 500 ps MD simulation discussed above. We saw that the anode and cathode carried net positive and negative charges, respectively, after the voltage was applied across the initial relaxed structure (Figure\,\ref{fgr:chargedistribution}\,(a)), as expected. Consequently, the electrodes would have repelled (attracted) the ions in the functional layers that had charges of the same (opposite) polarities. The resultant movement of the ions, as discussed in the previous section, led to a redistribution of the net charges within the functional layer after the 500 ps of MD simulation under the applied voltage, as shown in Figure\,\ref{fgr:chargedistribution}\,(b), minimizing the total electrostatic energy. Notably, a thin top region of the dielectric at the interface with the anode, which is above the 50~\AA  z position, carries a significantly high net negative charge. This can be attributed to this region in the dielectric mainly consisting of the negatively charged O ions or anions and being nearly devoid of any cations (Figure\,\ref{fgr:ionprofiles}\,(a and b)) - as a result of the Ta ions being pushed away (downwards) from the anode interface (Figure\,\ref{fgr:ionmotion300K}\,(a)). On the other hand, the increase in the net positive charge close to the cathode (which is below the 0~\AA z position) in the dielectric layers (Figure\,\ref{fgr:chargedistribution}\,(b)) is smaller in magnitude as compared to the increase in the net negative charge near the anode. This could be attributed to the fact that this increase is due to the higher accumulation of Hf ions in the dielectric layer near the cathode while, in this case, the concentration of anions remains fairly consistent in the same region (Figure\,\ref{fgr:ionprofiles}\,(a and d)). 

The reasons as to why the cations underwent significant voltage-driven displacement within the dielectric, including being repelled away from the anode, while, in contrast, the O anions did not get affected as strongly, could be understood by taking a closer look at the mean per atom charges on each ionic species. From Figures \ref{fgr:chargedistribution}\,(c and d), we note that the magnitude of atomic charges on the metal ions remains significantly higher than that on the O ions, with the charges on the Ta ions being more than double those on the O ions (as can be seen for the plots corresponding to both the initial and final structures in Figures \ref{fgr:chargedistribution}\,(c and d)). Consequently, the electrostatic forces on the metal cations due to the charges on the electrodes are much stronger as compared to those on the O anions, resulting in the observed characteristically different voltage-driven response of each ionic species (Figure\,\ref{fgr:ionprofiles} and Figure\,\ref{fgr:ionmotion300K}\,(a)). Interestingly, one can note from Figure \ref{fgr:chargedistribution}\,(d) that in the final structure, the mean charges on the oxygen atoms in the thin dielectric region close to the anode (top electrode) got reduced to almost 0. This aligns well with the expectation based on our previous observations that the thin region on top of the dielectric layers near the anode is almost devoid of any metal ions (Figure \ref{fgr:ionprofiles} (b)), as the ionic charges are primarily determined by their chemical environment \cite{xiao2019comparative}. 

In fact, such a redistribution of ions at the electrode interfaces carrying net charges of opposing polarities compared to that in the corresponding adjacent electrodes (Figure \ref{fgr:chargedistribution}~(b)) in response to the applied voltage should effectively shield the ions within the dielectric matrix from the majority of the external electric field. This could explain the need to apply a particularly high peak voltage of over 4\,V experimentally \cite{stecconi2022filamentary}, which can overcome this shielding effect, in order to electroform these TaO$_{\rm x}$/\ce{HfO2}-based ReRAM devices. This shielding of the bulk ions from the external electric field also subdues the ability of the applied voltage on the electrodes to cause continuous electroreduction of the dielectric and consequent migration of oxygen vacancies that can then contribute to filament growth, which is one of the key processes in the VCM model. 

\subsection{Effect of Joule Heating on the Atomistic Forming Mechanisms}

\begin{figure}[h!]
  \includegraphics[width=1\textwidth]{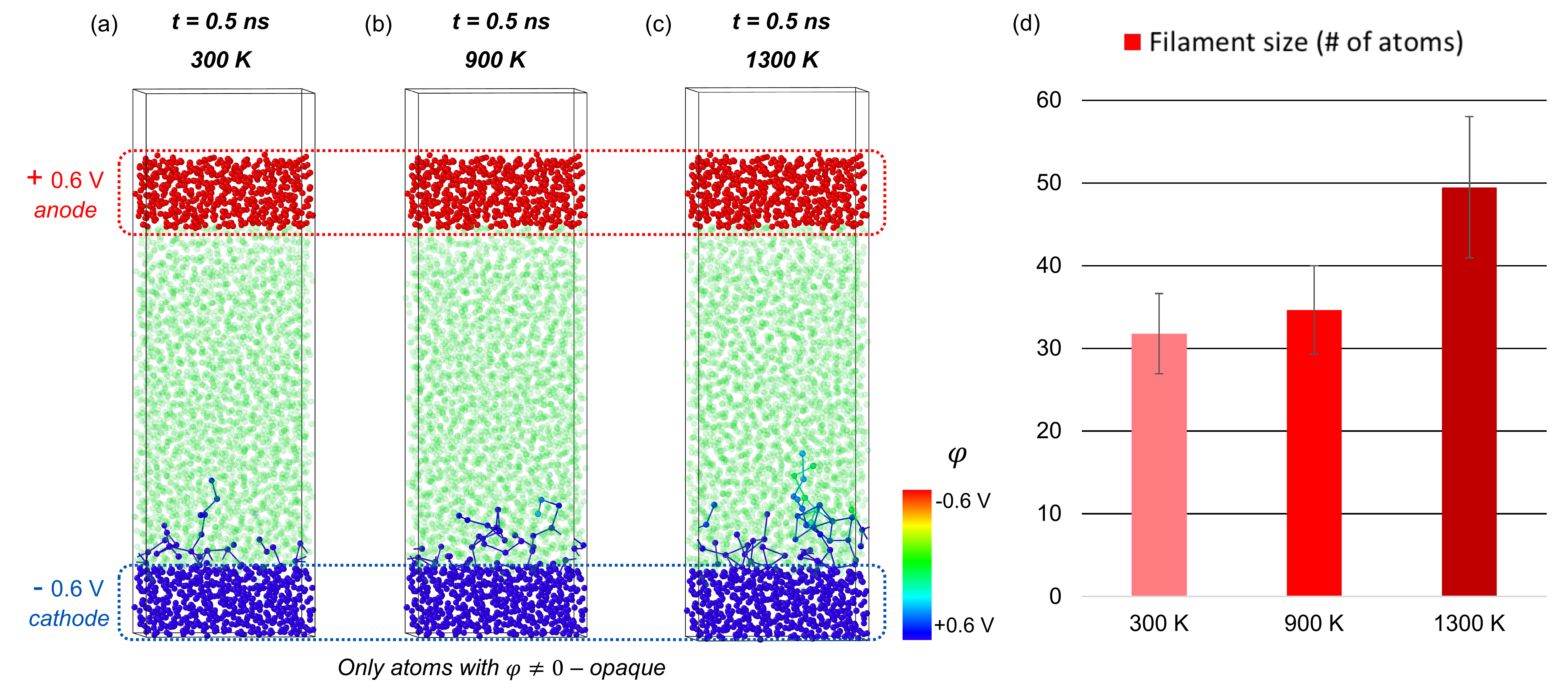}
  \caption{Effect of temperature on the growth of filament after 500\,ps simulation under 1.2\,V difference across the electrodes. Atomic snapshots of the final device structures post simulations at (a) 300 K, (b) 900 K, and (c) 1300 K, respectively. All the atoms with a 0 V electrochemical potential have been made semi-transparent. Consequently, only the electrode atoms and conductive cations that form the filament at the cathode interface are colored opaque. This is because the electrochemical potential applied to the cathode propagates through the conductive filament atoms, as indicated by the color coding. (d) The number of atoms in the filament for each temperature, averaged over the final 100\,ps of simulation time with the error bars showing the corresponding standard deviations.}
  \label{fgr:filament1.2V}
\end{figure}

To investigate the mechanism of filament formation and growth under Joule Heating in these devices, we repeated the simulation of our device model at two different temperatures of 900\,K and 1300\,K under the same voltage condition of 1.2\,V across the electrodes. These higher temperatures were selected based on two reported values of maximum internal temperatures at the onset of forming in two different \ce{TaO2}-based devices, respectively, that also had inert electrodes \cite{ma2020exchange}. Although the overall composition profiles, including the oxygen-rich region near the anode, for these two higher temperatures were similar to that for the earlier simulation at room temperature at the end of the simulations run for 500 ps (Figure\,\ref{fgr:structurechange}(b)), there were notable differences between the extents of filament growth at each temperature. Figure\,\ref{fgr:filament1.2V} shows the final atomistic snapshots (with the atoms of the electrodes and those forming the filaments color-coded to distinguish them from the rest of the dielectric matrix) and filament sizes, calculated as the average number of metallic ions forming the filament, for all the three simulations (at 300\,K, 900\,K, and 1300\,K) performed under an applied voltage of 1.2 V. These results indicated that the size of the nucleated filament increases significantly with temperature, especially at the high temperatures of 1300\,K in the functional layers (Figure~\ref{fgr:filament1.2V}\,(d)). 

\begin{figure}[h!]
  \includegraphics[width=0.8\textwidth]{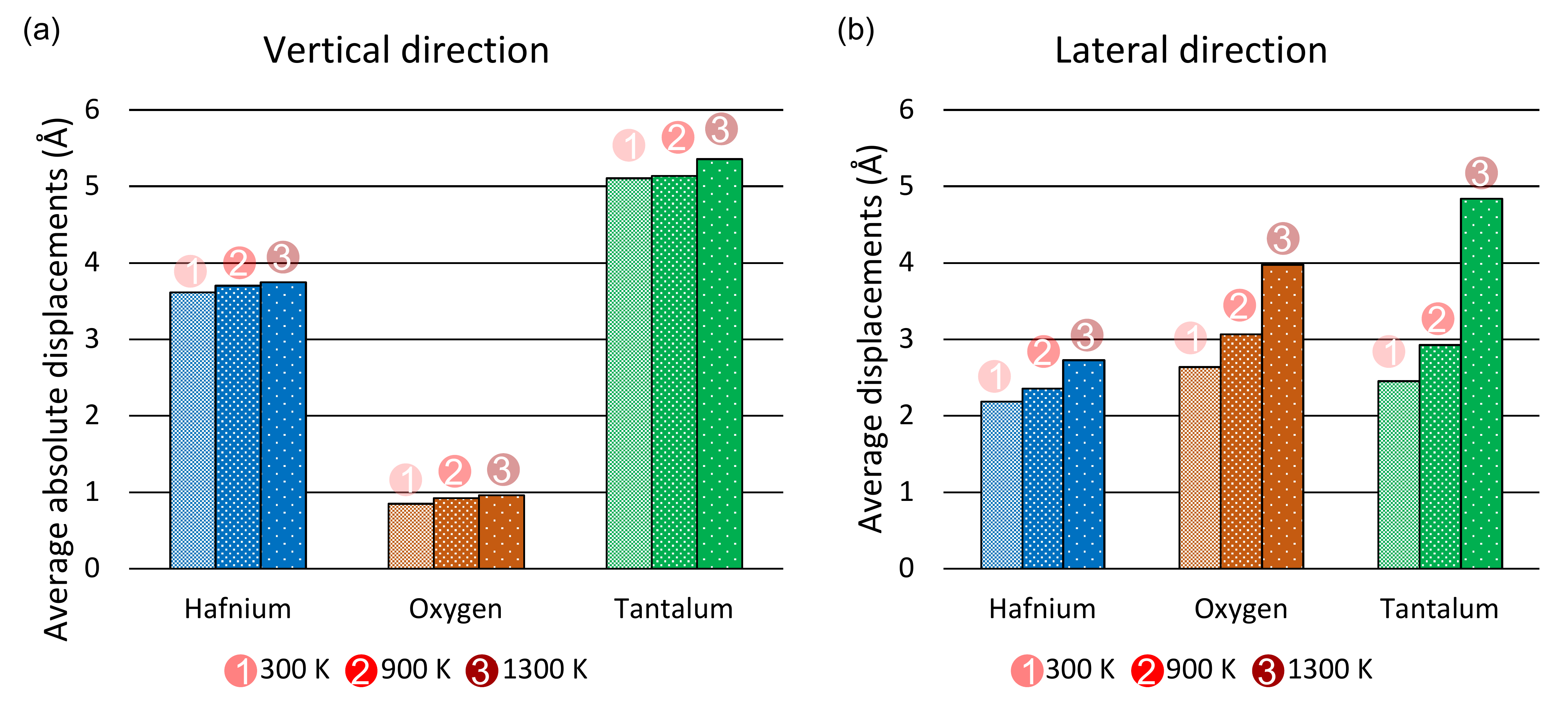}
  \caption{The change in average displacements of different ions for varying temperatures over 500\,ps of MD simulation under 1.2\,V difference across the electrodes. The average absolute displacements along the (a) vertical and (b) lateral directions in the device stack models.}
  \label{fgr:ionicmotion1.2V}
\end{figure}

In order to have a deeper understanding of the atomic processes underlying the filament growth at varying temperatures, we looked at the average net displacement of each atom type for the three 500 ps simulations, plotted in Figure\,\ref{fgr:ionicmotion1.2V}. Interestingly, the absolute displacement of each ionic species in the vertical direction, which is parallel to the external electric field, increased only by a small amount on increasing the temperature (Figure\,\ref{fgr:ionicmotion1.2V}\,(a)). These results indicate that the increase in the size of the filament at higher temperatures cannot be primarily attributed to thermally activated increase in the vertical defect migration along the direction of the external electric field. This is in contrast to what is understood in the VCM model\,\cite{dittmann2021nanoionic}. 

On the other hand, the lateral motion of the ions, which is predominantly thermally activated, increased significantly with temperature, with a notably high increase at 1300\,K for the O and Ta ions (Figure\,\ref{fgr:ionicmotion1.2V}\,(b)). In fact, at 1300\,K, the lateral motion of the Ta ions became even higher than that of the O ions, while the latter dominated at the temperatures of 300\,K and 900\,K (Figure\,\ref{fgr:ionicmotion1.2V}\,(b)). It is reasonable to assume that the high thermally activated motion of the ions (Figure\,\ref{fgr:ionicmotion1.2V}\,(b)) corresponds to an increased rate of bond breakage and O vacancy defect generation in the functional layers of the device, getting notably more pronounced at elevated temperatures (1300\,K in our simulations). The generated O vacancies near the top end of the pre-seeded filament get stabilized by the negative charges from the filament \cite{zeumault2021tcad}, which acts as a virtual cathode during forming \cite{dittmann2021nanoionic}. These nearby newly generated O vacancies should then agglomerate and combine with the existing filament, driven by the localized electric fields\,\cite{urquiza2021atomistic}, growing it in size (Figure\,\ref{fgr:filament1.2V}\,(a-c)). These results further corroborate and elucidate the key role of thermal effects in accelerating filament formation through a mechanism of thermally-driven increased defect generation in the vicinity of the growing filament. One of the consequences of this is that if we limit the peak forming temperatures through a compliance current set-up or a series resistor, then the joule heating is physically limited, which is very effective in lowering the conductance of the post-forming ReRAM due to a smaller size of the filament.

In fact, this mechanism that gets significantly more pronounced when the temperature reaches 1300\,K in our simulations (Figures\,\ref{fgr:filament1.2V} (b) and \ref{fgr:ionicmotion1.2V} (b)) can also explain the compositional runaway effect observed in other TaO$_{\rm x}$ and \ce{HfO2}-based ReRAMs and the associated abrupt acceleration of filament growth during the electroforming process when the local temperature inside the device gets high enough \cite{ma2020exchange, ma2018formation, kumar2016conduction, kumar2016memristors}.

Based on this understanding that the internal local temperature is the key factor underlying forming of the TaO$_{\rm x}$/\ce{HfO2}-based ReRAM, we can expect that increasing the overall temperature of the device should reduce the required forming voltage. This is because an already high device temperature would reduce the current density requirement for reaching the necessary internal temperatures in the functional layer by Joule heating to cause sufficient vacancy generation for rapid filament growth. We have also experimentally observed this when heating similar bilayer metal oxide-based ReRAM devices during electroforming, and found a strong linear correlation between the temperature of the device and the required peak electroforming voltage, as shown in Figure S2 of the SI . 

\subsection{Atomistic response to 0.6 V applied across pristine device stack}

\begin{figure}[h!]
  \includegraphics[width=1\textwidth]{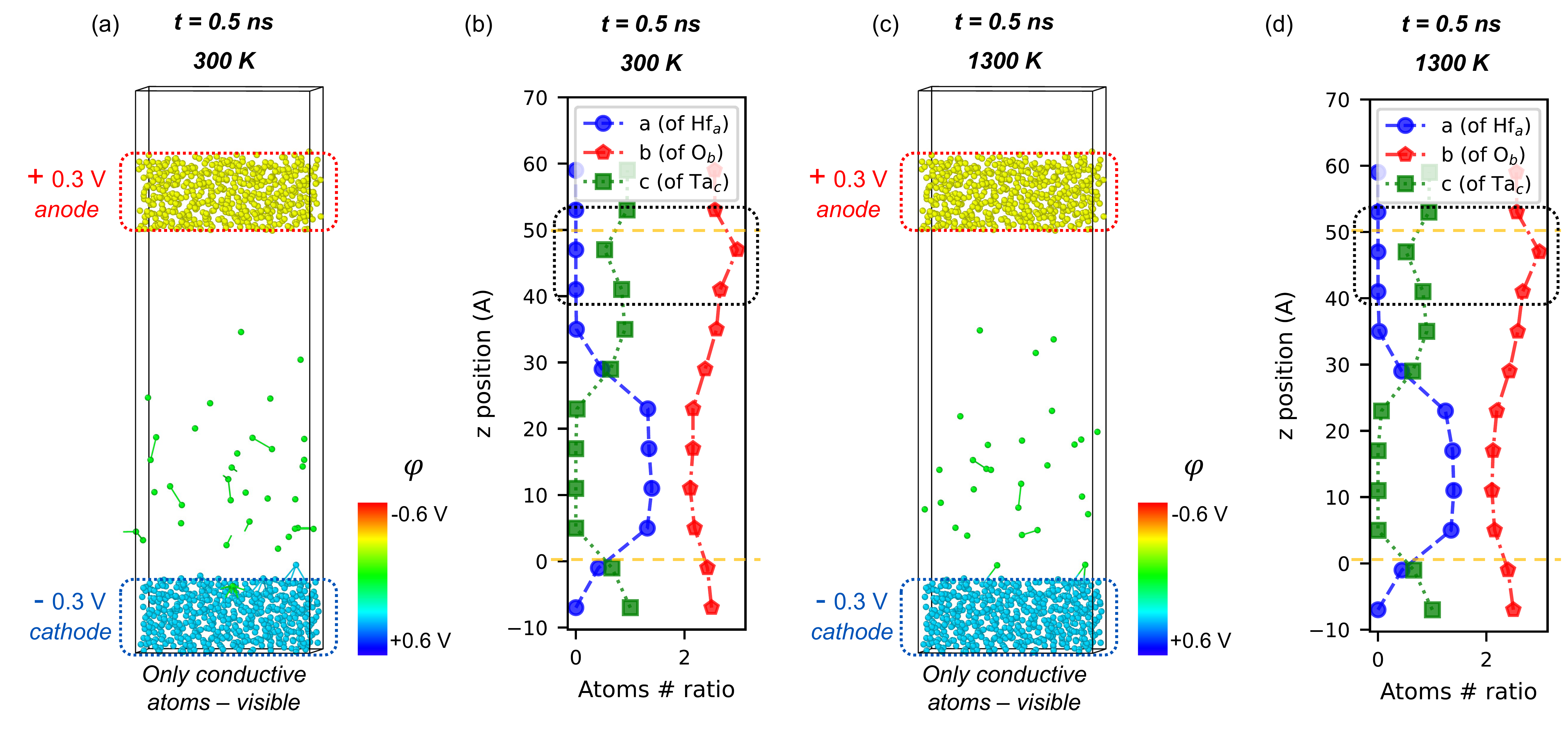}
  \caption{Effect of temperature on the vacancy distribution and composition change in the device stack after 500\,ps simulation with a 0.6\,V difference applied across the electrodes. (a) and (b) show the atomic snapshots of the device models simulated at 300\,K and 1300\,K, respectively. Only the electrode atoms and the conductive metal ions are visualized for clarity. (b) and (d) plot the profiles of the varying proportion of different ions within the device stack at the end of the corresponding simulations done at 300\,K and 1300\,K, respectively. The a, b, and c values, corresponding to the proportion of Hf, O, and Ta atoms, respectively, in the structure at any position along the z-axis, are normalized so that $a+b+c$ equals 3.5. The upper and lower dotted yellow lines in each subplot mark the location of the interfaces of the functional layers with the anode and the cathode, respectively, in the device model.}
  \label{fgr:change0.6V}
\end{figure}

We next simulated our device model for 500 ps under 0.6 V applied across the electrodes, first at 300 K and then at the elevated temperature of 1300 K. This voltage was found to be low enough that the filament nucleation response, observed earlier under the higher 1.2 V external bias (Figure \ref{fgr:structurechange}~(a)), did not occur (as elaborated further below) and, thus, allowed us to study the associated atomistic mechanisms and the thermal effects under a voltage that is well below the threshold needed for the onset of forming. The state of vacancy distribution in the final structures and the final composition profiles are shown in Figure \ref{fgr:change0.6V}. We note that at 300 K, the vacancies did not agglomerate to initialize the filament formation in the device (Figure \ref{fgr:change0.6V}~(a)), while an oxygen-rich region still developed near the anode (Figure \ref{fgr:change0.6V}~(a-b)). We observe similar results at 1300 K as well under the 0.6 V applied bias (Figure \ref{fgr:change0.6V} (c-d)), i.e., the vacancies still could not cluster together to start filament formation even at the elevated temperature (Figure \ref{fgr:change0.6V} (c)).

\begin{figure}[h!]
  \includegraphics[width=0.8\textwidth]{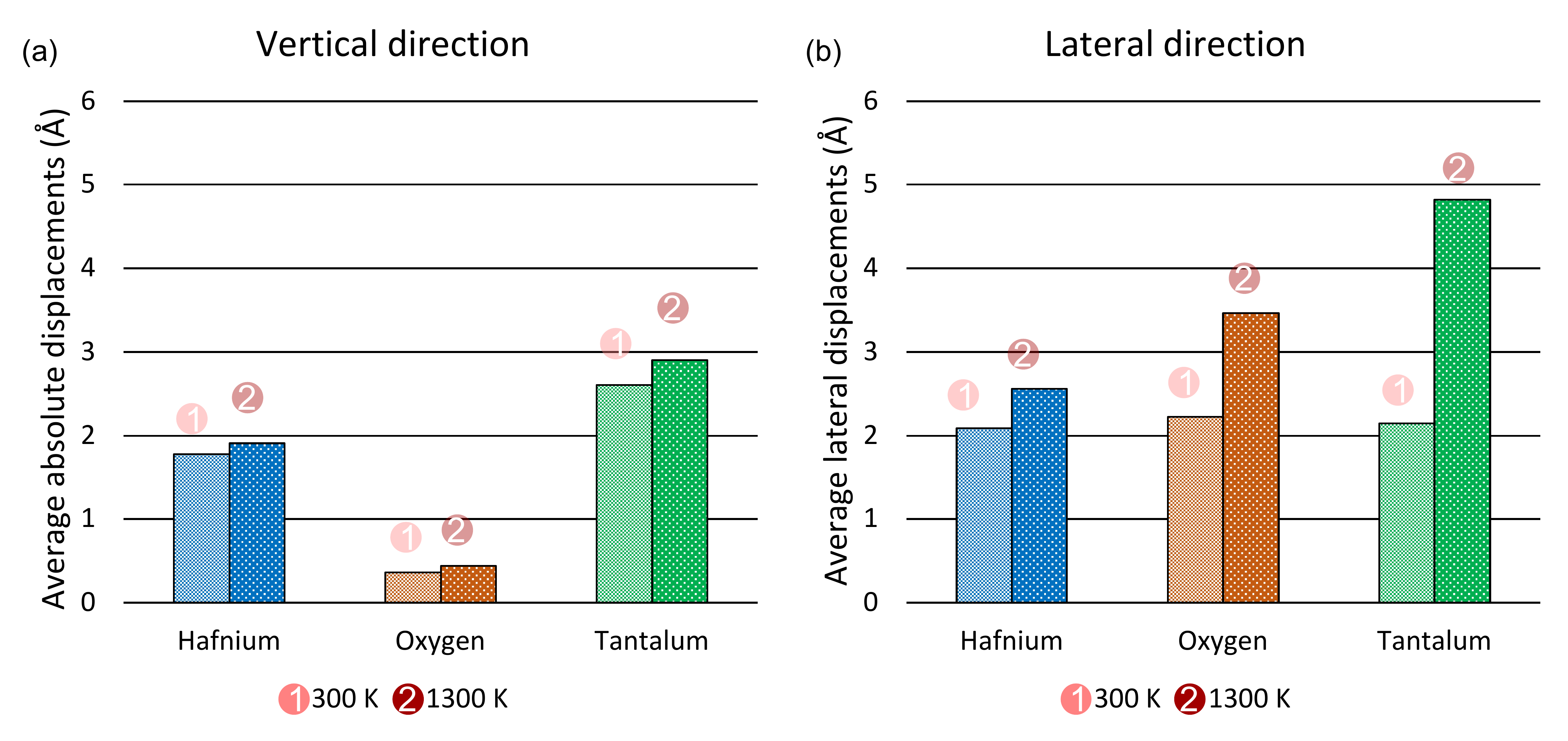}
  \caption{The change in average displacements of different ions for different temperatures over 500 ps of MD simulation under 0.6 V difference applied across the electrodes. The average absolute displacements along the (a) vertical and (b) lateral directions in the device stack models.}
  \label{fgr:ionicmotion0.6V}
\end{figure}

We further examined the ionic motion under a 0.6 V bias, as shown in Figure~\ref{fgr:ionicmotion0.6V}. We observed that increasing the temperature to 1300 K affects the ionic mobilities in the same way as it did when a higher 1.2 V bias is applied (Figure \ref{fgr:ionicmotion1.2V}). Specifically, the increased temperature significantly enhances the lateral motion of the ions, primarily for Ta, followed by O ions. However, despite this similar increase in mobility on increasing the temperature, which was attributed as the cause for accelerating the filament growth under the 1.2\,V applied voltage, no filament nucleation occurs under the low bias condition. This suggests that the applied voltage plays a critical role in the electroforming process beyond just causing the drift of vacancy defects towards the cathode. The external voltage helps overcome the repulsion between vacancies, allowing them to cluster together forming the connected network of conductive phases, or the filament. 

This highlights an important finding that the formation of a filament in TaO$_{\rm x}$/\ce{HfO2}-based ReRAMs can only occur if the applied voltage exceeds a certain threshold that is needed for the vacancy clusters to start agglomerating. Therefore, while thermal effects can accelerate filament growth, they only have an impact when the applied voltage surpasses this threshold in a given bilayer ReRAM device, enabling the filament to nucleate in the first place.

\begin{figure}[h!]
  \includegraphics[width=0.5\textwidth]{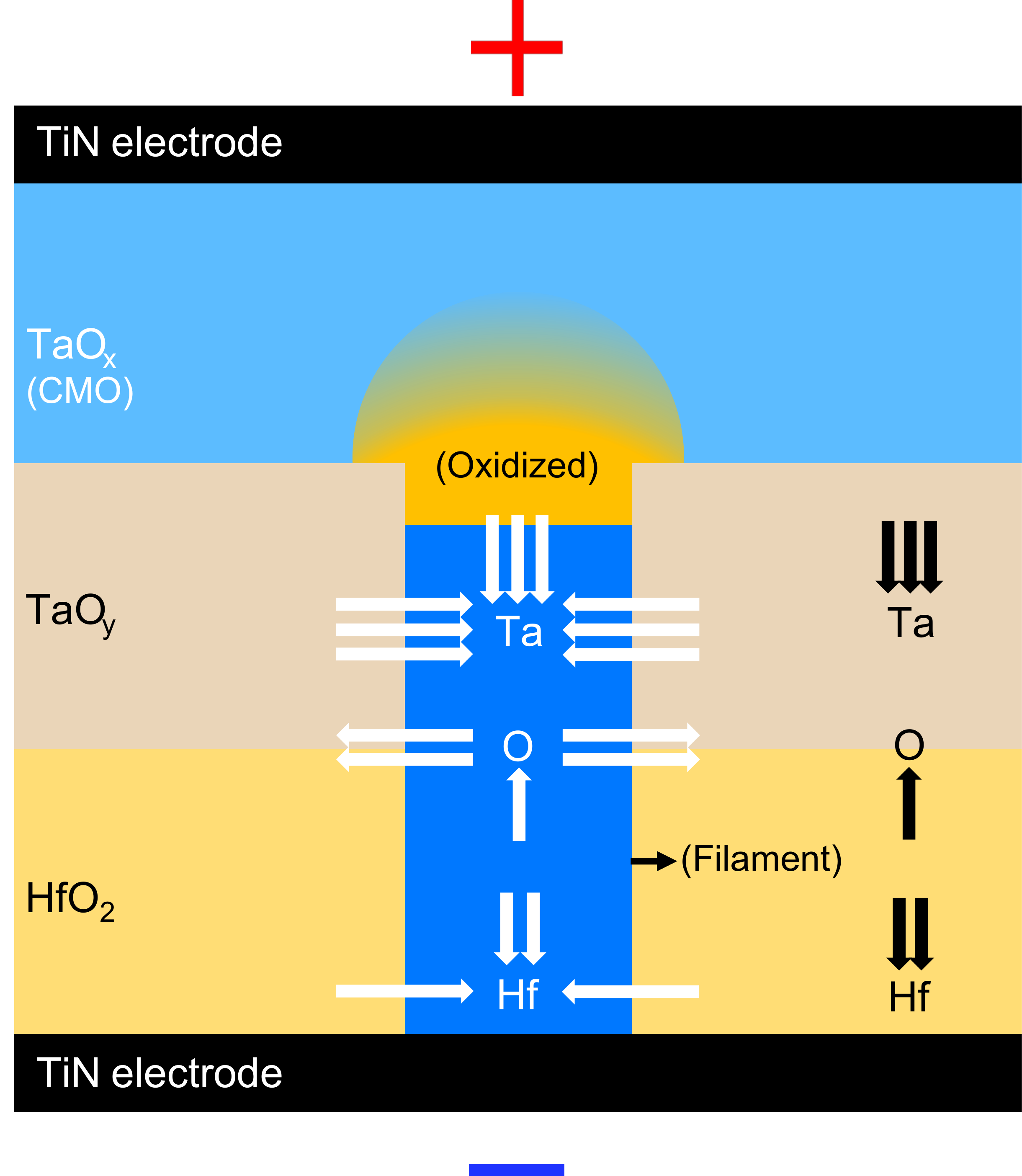}
  \caption{Schematic illustrating the possible atomic movements underlying forming in TaO$_{\rm x}$/HfO$_{\rm 2}$ ReRAM devices. The number of arrows qualitatively represents the relative magnitudes of displacement of each ion in different directions and across different regions within the device.}
  \label{fgr:schematic}
\end{figure}
In this study, we revealed the detailed atomistic forming mechanisms in the bilayer TaO$_{\rm x}$/HfO$_{\rm 2}$-based ReRAM and clarified their physical underpinnings. These insights are illustrated in Figure \ref{fgr:schematic}, which includes a schematic of a possible final structure of the formed ReRAM device. The atomic movements behind the forming are qualitatively represented by arrows, with their number indicating the relative magnitudes of displacement for each ion in different directions and regions within the device.

The temperature and voltage-driven motion of the different ionic species as observed in this study for both the functional layers indicates that the filament formation occurs across both the functional dielectric layers in the TaO$_{\rm x}$/HfO$_{\rm 2}$-based ReRAM devices, along with the creation of an oxygen-rich region at the interface between the conductive TaO$_{\rm x}$ and dielectric TaO$_{\rm y}$ layers. The extension of the oxidized region into the TaO$_{\rm x}$ layer on top of the filament, as proposed in prior experimental studies \cite{stecconi2022filamentary}, could occur at the high temperatures reached during forming as a result of Fick's diffusion of oxygen ions accumulated at the interface \cite{ma2020exchange}.

\section{Discussion and Conclusions}

We performed reactive all-atom MD simulations on a TaO$_{\rm x}$/HfO$_{\rm 2}$-based bilayer resistive switching device by locally applying voltage to inert electrodes, using an extended CTIP formalism tailored for multi-metal oxide systems in combination with the EChemDID method. Upon applying a 1.2 V bias, to investigate early-stage electroforming, to the pristine device model, we observed the nucleation of a filament at the interface with the negative bottom electrode (cathode). Simultaneously, an oxygen-rich region developed near the anodic interface-corresponding to the conductive top layer acting as an extension of the top electrode; within the Ta$_{\rm 2}$O$_{\rm 5}$ dielectric interlayer. This was driven by the repulsion of Ta atoms away from the anode. Notably, both Ta and Hf ions exhibited significantly larger vertical displacements than oxygen ions along the direction of the applied electric field, which we attribute to their higher per-atom charges. Our observations align well and can explain the experimental observations on some tantalum and hafnium oxide-based ReRAMs that indicate that the mobile species in these devices are primarily the metal cations \cite{wedig2016nanoscale,ma2019stable}.

On increasing the temperature up to 1300 K, we observed a significant increase in the size of the nucleated filament, in terms of the number of metallic ions that are part of the filament at the end of the simulation. This was accompanied by significantly enhanced lateral displacements of the ions, particularly of O and Ta ions, and only a minor increase in the vertical voltage-driven ionic migration. Thus, the role of temperature in accelerating filament growth was attributed to a thermally activated increase in the generation of O vacancy defects. These defects get stabilized in the vicinity of the growing filament, which acts as the virtual cathode, and aggregate with it to cause rapid filament growth. Indeed, studies on monoclinic-\ce{HfO2} have shown that the energy barrier for the generation of O vacancy-interstitial defect pairs remains significantly high (over 6 eV) even when the electric fields are close to the breakdown limit, whereas the injection of extra electrons from the cathode can significantly lower this energy barrier to around 1 eV \cite{strand2020effect}. Moreover, our results further help to elucidate the mechanisms underlying the experimental observation of predominant lateral ionic motion accompanying forming in tantalum and hafnium oxide-based single layer ReRAM devices as well \cite{kumar2016conduction,kumar2016memristors,ma2020exchange}.

We also found the existence of a threshold voltage beyond which the oxygen vacancies in the functional layer can overcome their mutual repulsion and cluster together near the cathode, enabling filament nucleation. This coupled temperature and electric field-driven mechanism can explain the previous reports of these devices exhibiting behaviors characteristic of both the VCM and TCM models\,\cite{ma2020exchange, ma2019stable,goodwill2017electro,kumar2016conduction,kumar2016memristors,ma2018formation}. We also experimentally observed that increasing the chip temperature reduced the forming voltage of a metal oxide-based bilayer ReRAM device, which aligned well with the role of the thermally-driven filament-forming mechanisms in these devices as deduced from our simulations.

Thus, the mechanistic insights from this study provide a clear atomic-level understanding of the various processes underlying the forming behavior of the TaO$_{\rm x}$/\ce{HfO2}-based ReRAM devices. Notably, we found that the local electric field and Joule heating in the device combine to accelerate filament growth during forming through an interfacial mechanism of defect generation, stabilization, and agglomeration. This alleviates the requirement of continuous anodic electro-reduction of the dielectric and the subsequent electro-migration of the oxygen vacancies for electroforming, which is part of the VCM model\,\cite{waser2009redox,padovani2012understanding,jeong2008characteristic,urquiza2021atomistic,dittmann2021nanoionic}. These insights are expected to not only guide future optimization and deliberate improvements in the device design, but also the development of more physics-accurate analytical and compact models for emulating the behavior of these devices for circuit-level simulations.

In this study, we extended the CTIP + EChemDID formulation and implemented it as an additional package for the open-source LAMMPS simulation tool\,\cite{plimpton1995fast}, which can aide future all-atom reactive MD simulations of multi-metal oxide systems under the application of an external voltage bias. It is important to note that the complete electroforming process typically occurs in the timescale of microseconds\cite{ma2020exchange}, which exceeds the timescales accessible by all-atom reactive MD simulations. Therefore, our analysis focuses on the voltage- and temperature-driven mechanisms that govern the early stages of filament formation and growth in the bilayer ReRAM device. While we can combine these insights to understand the underlying physics of the full forming process in this device, further work is required to draw definitive conclusions about the structure of the device at the end of the forming process, which is needed to verify the possible switching mechanisms in TaO$_{\rm x}$/\ce{HfO2}-based ReRAM devices.

\section{Methodology}
\label{sec:methods}

\subsection{Interatomic Potential and Charge Dynamics}

Atomic interactions were modeled using the CTIP formalism, which is a kind of hybrid reactive potential proposed by Zhou \textit{et al.} that can model both the metallic and ionic interactions in a single formalism \cite{zhou2004modified,zhou2005charge}. Coupled with the charge equilibration (QEq) scheme, it considers a radial distribution of electron densities instead of point charges and allows dynamic variation of atomic charges \cite{rappe1991charge,lee2016modified,wu2023developing}. CTIP is a modification of the original Streitz and Mintmire potential \cite{streitz1994electrostatic} with added charge bounds, which overcomes several of its limitations and has been optimized to accurately model the redox reactions in a system comprising multiple metal oxides, including HfO$_{\rm 2}$ and TaO$_{\rm x}$ \cite{sasikumar2017evolutionary,sasikumar2019machine,wu2023developing}. In this modified CTIP formalism, the total energy is partitioned as a combination of the electrostatic energy with the embedded atom method (EAM) for non-electrostatic contributions\,\cite{zhou2004modified,zhou2005charge,wu2023developing} as shown below

\begin{equation}
\label{eq:ctip}
    E = E_{es} + E_{ne}.
\end{equation}
%In Equation~\ref{eq:ctip},
$E_{es}$ is the electrostatic contribution given as\,\cite{wu2023developing}

\begin{eqnarray}
\label{eq:Ees}
%\begin{split}
    E_{es} &=& E_{0} + \sum_{i=1}^{N} q_{i}\chi_{i}^0 + \frac{1}{2}\sum_{i=1}^{N}\sum_{j=1}^{N}q_{i}q_{j}V_{ij} + \sum_{i=1}^{N}\omega\left( 1-\frac{q_{i}-q_{min,i}}{|q_{i}-q_{min,i}|} \right)(q_{i}-q_{min,i})^2 \nonumber\\
    &+& \sum_{i=1}^{N}\omega\left( 1-\frac{q_{max,i}-q_{i}}{|q_{max,i}-q_{i}|} \right)(q_{max,i}-q_{i})^2,
%\end{split}
\end{eqnarray}
%\end{equation}
%In Equation \ref{eq:Ees}, 
wherein $N$ represents the number of atoms in the system, $q_{min,i}$ and $q_{max,i}$ represent the bounds for the charges on an atom $i$, which is imposed by an energy penalty when the charge is beyond these bounds. The strictness of the bounds is determined by the coefficient $\omega$\,\cite{zhou2004modified, zhou2005charge}. $\chi_{i}^0$ and $V_{ij}$ represent the electronegativity of the atom and the electrostatic potential contribution arising from the interaction between partial charges on atoms $i$ and $j$, respectively. The values of the parameters for this equation, which were optimized for the corresponding multi-metal oxide systems, have been obtained from the literature\,\cite{wu2023developing,sasikumar2017evolutionary,sasikumar2019machine,zhou2004modified}.

The non-electrostatic contribution, $E_{ne}$, described by Zhou and Wadley\,\cite{zhou2004modified} is given as
\begin{equation}
\label{eq:Ene}
\begin{split}
    E_{ne} = \frac{1}{2} \sum_{i=1}^{N }\sum_{j=i_1}^{i_M}\varphi_{ij}(r_{ij}) + \sum_{i=1}^{N}F_{i}(\rho_{i}),
\end{split}
\end{equation}
The first term accounts for the pairwise interaction energy between atom $i$ and its neighboring atom $j$, where $\varphi_{ij}(r_{ij})$ represents the interaction potential as a function of the interatomic distance $r_{ij}$. The summation over $j \in N_i$ includes all the neighboring atoms of atom $i$. The second term in Equation \ref{eq:Ene} represents the embedding energy $F_{i}(\rho_{i})$, which quantifies the energy required to place atom $i$ into the local electron density $\rho_{i}$ generated by its surrounding atoms. The value tables for $\rho_{i}$, $F_{i}(\rho_{i})$, and $\varphi_{ij}(r_{ij})$ for different interatomic separation distances have been calculated using suitable functional forms and parameters optimized for the corresponding multi-metal oxide systems in literature\,\cite{wu2023developing,sasikumar2017evolutionary,sasikumar2019machine,zhou2004modified}. The atomic charge equilibration was achieved through the electronegativity equalization method (EEM)\,\cite{mortier1986electronegativity}.

\subsection{Generation of Amorphous Atomistic Structure Models}

The amorphous \ce{HfO2} and \ce{Ta2O5} structures were generated using a melt-and-quench molecular dynamics protocol\,\cite{urquiza2021atomistic, xiao2019comparative}. A 5 × 5 × 6 supercell of monoclinic \ce{HfO2} and a 7 × 6 x 5 supercell of triclinic \ce{Ta2O5} were taken as the corresponding initial atomistic models. The cohesive energy values obtained for the energy minimized crystalline \ce{HfO2} and \ce{Ta2O5} models were 7.79 eV/atom and 7.03 eV/atom, respectively, with corresponding densities of 10.19 g/cm$^3$ and 8.39 g/cm$^3$, respectively, which agree well with previously calculated values reported in literature using the same interatomic potential\,\cite{wu2023developing} providing a validation for our forcefield implementation. These energy minimized initial crystalline structure models were heated to 3300 K and 5000 K for hafnia and tantala, respectively, and equilibrated at this temperature for 100 ps to ensure complete melting. The liquid phase was quenched to 1 K over 100 ps and 150 ps intervals, respectively, for the hafnia and tantala models. Throughout the simulation, the system was maintained at 1 bar pressure by allowing volume fluctuations along the z-axis, while the lateral dimensions were constrained to 2.5 × 2.5 nm$^2$ to remain consistent across all the layers of the final device stack model. The radial distribution functions and the distribution of M-O coordination numbers with a cut-off radius of 2.7\,\AA\, calculated for the final amorphous structure models (see Figures S3 and S4 of the SI ) agreed well with the literature \cite{urquiza2021atomistic, xiao2019comparative, sasikumar2017evolutionary, sasikumar2019machine, wu2023developing}.

\subsection{External Electrochemical Potential Propagation Method}

To simulate bias-induced redox reactions and filament formation in the system, the EChemDID method \cite{onofrio2015voltage} was extended to work together with the CTIP formalism and implemented in LAMMPS. In this method, an external voltage, $\Phi_0$, is applied by adding a suitable value of the local potential ($\Phi$) to the intrinsic electronegativities of electrode atoms as $\chi_i^0 \to \chi_i^0 - \Phi_0/2$ for the anode or the positive electrode (i.e., $\Phi$ = $-\Phi_0/2$), and $\chi_i^0 \to \chi_i^0 + \Phi_0/2$ for the cathode or the negative electrode (i.e., $\Phi$ = $\Phi_0/2$) \cite{onofrio2015voltage}. The metallic atoms in the model at any instant were identified as the Ta and Hf atoms with oxygen coordination number $\leq$ 5 following a previous study~\cite{urquiza2021atomistic}. The connected networks of electrode and metallic atoms that form the filaments within the structure were identified through cluster analysis after every 50 ps of MD simulation. After this step, the local potential values for the atoms identified as part of the filaments were calculated by solving a computationally efficient nonphysical diffusion equation\,\cite{urquiza2021atomistic,onofrio2015voltage,onofrio2015atomic} given below, using the atoms as a grid:

\begin{equation}
\label{eq:iondiffusion}
\begin{split}
    \dot\Phi = k \sum_{j\neq i} \frac{\Phi_i(t) - \Phi_j(t)}{|R_{ij}|^2}w(R_{ij}),
\end{split}
\end{equation}
%This equation 
In Equation \ref{eq:iondiffusion},  $k$ is an effective diffusivity and $w(R_{ij})$ is a weight function calculated as:
\begin{equation}
\label{eq:weight}
w(R_{ij})=
\begin{cases}
    N\left[ 1 - \left( \frac{R_{ij}}{R_C} \right)^2 \right]^2 & \text{if } R_{ij} < R_C \\
      0 & \text{otherwise }
\end{cases}
\end{equation}
wherein $N$ is a normalization constant, taken as 0.5, and $R_C$ is the cutoff radius, taken as 3.9\,\AA, for two metallic atoms to be part of the same cluster\,\cite{urquiza2021atomistic,onofrio2015voltage}. The relaxation term in the originally proposed diffusion equation of the EChemDID method was dropped in Equation \ref{eq:iondiffusion} \cite{onofrio2015atomic,urquiza2021atomistic}. Instead, the local potential values for the atoms not part of either the filaments or the electrodes were assigned as 0\,V after each update of the filament, reducing the computational complexity without compromising the accuracy in simulating the filament evolution process \cite{urquiza2021atomistic}. 

The propagation of the local potential through the metallic atoms was observed and used to validate the implementation of this method (depicted in Figure S5 of the SI) and can be optimized by changing $k$ and the duration of the voltage equilibration process. We note that the exact dynamics of the voltage equilibration process were not critical in our simulations as long as the charge distributions as a result of the propagation of the local potential through the metallic atoms in the structure at any instant were produced self-consistently\cite{onofrio2015voltage}. In our simulations, the diffusive step of Equation \ref{eq:iondiffusion} was run for 10 iterations per MD step, after which the resultant per atom local potential values were added to the atomic electronegativities used for atomic charge equilibration\,\cite{mortier1986electronegativity}. This was further repeated for 0.2\,ps without any dynamics followed by MD simulations for the next 49.8\,ps until the next filament identification and update. This helps to capture the evolving nature of the filaments by simulating the ionic dynamics and redox reactions together with the varying charges and localized electric fields within the bilayer device model self-consistently during the simulation. A timestep of 0.2 fs was adopted for the MD simulations.

%%%%%%%%%%%%%%%%%%%%%%%%%%
\section{Supporting Information} %%%See SI.tex
Figures~S1 $\to$ S5 are provided that include STEM-EDS line profile for the device cross section, experimental peak forming voltage versus device temperature, structural information for the amorphous models, and EChemDID voltage equilibration test results.

The extended CTIP-EChemDID package developed as part of this work for LAMMPS, along with an explanation for compilation, example simulation, and analysis scripts, will be available and maintained by the authors at this \href{https://github.com/simantalahkar/LAMMPS-CTIP-EChemDID}{GitHub Repository}. Any other data related to this study will be provided upon reasonable request.

%%%%%%%%%%%%%%%%%%%%%%%%%%%%%%%%%%%%%%%%%%%%%%%%%%%%%%%%%%%%%%%%%%%%%
%% The "Acknowledgement section" can be given in all manuscript
\begin{acknowledgement}

The authors thank the European Union's Horizon Europe research and innovation program under PHASTRAC project grant agreement No. 101092096 for funding this study.
The authors also thank SURF (www.surf.nl) for the support in using the National Supercomputer Snellius. This work used the Dutch national e-infrastructure with the support of the 
SURF Cooperative using grant no. EINF-11416.

\end{acknowledgement}

%%%%%%%%%%%%%%%%%%%%%%%%%%%%%%%%%%%%%%%%%%%%%%%%%%%%%%%%%%%%%%%%%%%%%
\bibliography{achem.bib}

%% graphical table of contents. It is given here since some journals
%%I comment the following out 
%\begin{tocentry}

%Some journals require a graphical entry for Table of Contents.
%This should be laid out ``print ready'' so that the sizing of the

%Inside the \texttt{tocentry} environment the font used is %Helvetica
%8\,pt, as required by \emph{Journal of the American Chemical

%The surrounding frame is 9\,cm by 3.5\,cm, which is the maximum
%permitted for  \emph{Journal of the American Chemical Society}
%graphical table of content entries. The box will not resize if the
%content is too big: instead it will overflow the edge of the box.

%This box and the associated title will always be printed on a
%separate page at the end of the document.

%\end{tocentry}

\end{document}